# Generic lepton mass matrices and neutrino oscillations

Rohit Verma

*Rayat Institute of Engineering and Information Technology, Ropar 140001, India*

*e-mail address:* rohitverma@live.com

Several intriguing aspects of neutrino oscillation phenomenology like the origin of small neutrino masses, the absolute neutrino mass scale, the neutrino mass hierarchy, i.e. normal or inverted and the nature of neutrinos, i.e. Dirac or Majorana, etc. have been addressed from a general perspective. We show how the fundamental considerations of unitary transformations, naturalness and seesaw mechanism suffices to determine the texture structure of fermion Yukawa couplings and discuss the significance of the effective mass in 0νββ decay for the texture structure of these couplings.

## INTRODUCTION

Unlike the quark sector, the recent measurement of large $\theta_{13}$ [1] along with indications of a non-maximal $\theta_{23}$ [2], point towards an absence of symmetry in the lepton sector [3], indicating that the masses and flavor mixing schemes for quarks and leptons are significantly different. This makes the task of constructing the corresponding mass matrices more challenging, especially if the fermion mass matrices are to be considered in a unified framework [4]. In the absence of a compelling theory of fermion flavor dynamics from the "top-down" perspective, several phenomenological "bottom-up" approaches have been widely adopted [5] including radiative mechanisms, texture zeros, family symmetries and seesaw mechanism, etc. Among these, the "texture zero" ansätze initiated by Weinberg and Fritzsch [6] has been quite successful in explaining the fermion masses and flavor mixing patterns. As a result, several texture zero possibilities of lepton mass matrices have been investigated in the literature [7].

Difficulty in implementing this approach arises from large redundancy in fermion mass matrices in SM framework, wherein the flavor structure of these matrices is not constrained by the gauge symmetry. As a result these matrices are arbitrary complex matrices involving 36 parameters, way large as compared to the number of physical observables. This redundancy is related to the fact that one has the freedom to make Weak Basis (WB) transformations under which the fermion mass matrices change but the gauge currents remain diagonal and real [8, 9].

An inspiring systematic study of these texture zeros originating from WB transformations was recently discussed in detail by Branco *et al.* [8] and also discussed earlier by Fritzsch *et al.* [9], wherein it emerged that some sets of these zeros have, by themselves, no physical meaning, since these can be obtained starting from arbitrary fermion mass matrices by making the appropriate unitary transformations, the so called WB transformations. This greatly reduces the number of possible sets of texture zeros to be studied. It was also observed [8,9] that using the freedom of such transformations, it is possible to obtain Hermitian fermion mass matrices involving a "maximum" of three phenomenological texture zeros "without" having any "physical" implications. Any "additional" texture zero is supposed to have physical implications.

Among the several possibilities of texture specific lepton mass matrices, the Hermitian Fritzsch-type lepton mass matrices involving six [7, 10] texture zeros have been exhaustively studied due to lesser number of parameters involved in these matrices. Such matrices were recently revisited by Fukugita *et al.* (FSTY) [10]. They observed that such matrices are able to explain the current neutrino oscillations data for the case of Majorana neutrinos exhibiting normal mass hierarchy (NH). However as these matrices involve greater number of texture zeros, they may not be considered as the most general lepton mass matrices in view of the WB transformations.

The purpose of the present paper is to use the fundamental considerations of unitary transformations, naturalness and seesaw mechanism to obtain lepton mass textures assuming normal neutrino mass hierarchy (NH). Following a hierarchical parameterization for these mass matrices, the exact relations for the lepton mixing angles have also been detailed, wherein the implications of neutrino mass hierarchy for neutrino oscillations are clearly evident. The significance of the effective mass involved in (0νββ) on the imposition of texture zeros in the resulting lepton mass matrices has also been explored.

## UNITARY TRANSFORMATIONS

In the WB approach [8,9], one may consider a basis wherein one of the lepton mass matrices among the charged lepton mass matrix $M_e$ and Dirac neutrino mass matrix $M_{\nu D}$ is real diagonal while the other is an arbitrary Hermitian mass matrix, for e.g.

$$M_e = D_e, \quad M_{\nu D} = V D_{\nu D} V^\dagger = \begin{pmatrix} e_{\nu D} & |a_{\nu D}|e^{i\alpha_{\nu D}} & |f_{\nu D}|e^{i\omega_{\nu D}} \\ |a_{\nu D}|e^{-i\alpha_{\nu D}} & d_{\nu D} & |b_{\nu D}|e^{i\beta_{\nu D}} \\ |f_{\nu D}|e^{-i\omega_{\nu D}} & |b_{\nu D}|e^{-i\beta_{\nu D}} & c_{\nu D} \end{pmatrix}. \quad (1)$$

Here $D_e = \text{diag}(m_e, -m_\mu, m_\tau)$ and $D_{\nu D} = \text{diag}(m_{\nu 1D}, -m_{\nu 2D}, m_{\nu 3D})$ are real diagonal matrices and V is the neutrino mixing matrix also called Pontecorvo-Maki-Nakagawa-Sakata (PMNS) matrix [11,12]. It has been shown [13] that for the quark sector, the observed hierarchy among the quark masses as well as the quark mixing matrix [12] elements get naturally translated onto the structure of the corresponding quark mass matrices i.e.

$$e < (|a|, |f|) < d < |b| < c. \quad (2)$$

Such hierarchical mass matrices have been referred to in the literature as *natural mass matrices* [14]. However, it shall be interesting to investigate the conditions under which the lepton mass matrices emerging through WB transformations exhibit such a structure, especially if neutrinos are Majorana type with the seesaw mechanism accounting for the small neutrino masses. In principle, an exact diagonalization of the mass matrix given in Eq. (1) is not always possible. In this context, one can apply a WB transformation [8, 9], which is essentially a unitary transformation U operating simultaneously on the mass matrices $M_e$ and $M_{\nu D}$, defined in eqn. (1) such that

$$M_e \to M'_e = U M_e U^\dagger, \quad M_{\nu D} \to M'_{\nu D} = U M_{\nu D} U^\dagger. \quad (3)$$

It is trivial to check that the two representations $(M_e, M_{\nu D})$ and $(M'_e, M'_{\nu D})$ are physically equivalent leading to the same mixing matrix, provided neutrinos are assumed to be Dirac particles. In such a case, as discussed in ref. [9], there is a possible choice of U such that

$$(M'_e)_{13, 31} = (M'_{\nu D})_{13, 31} = (M'_{\nu D})_{11} = 0, \quad (4)$$

or $\quad (M'_{\nu D})_{13, 31} = (M'_e)_{13, 31} = (M'_e)_{11} = 0, \quad (5)$

with non-vanishing other elements. However, in case the neutrinos are of Majorana type, the light neutrino mass matrix is obtained using the seesaw mechanism [15] as $M'_\nu = -M'^T_{\nu D} M_R^{-1} M'_{\nu D}$, where $M_R$ is the right handed Majorana mass matrix. It has been recently shown [10,16] that for Hermitian lepton mass matrices $M'_e$ and $M'_{\nu D}$ and for a real diagonal $M_R = m_R I$, where I is a unit matrix and $m_R$ denotes a very large mass scale, the *left* diagonalizing transformations for $M'_{\nu D}$ and $M'_\nu$ remain the same, as shown in Eqns. (9) and (11) below. Such a simple $M_R$ structure allows the two representations $(M'_e, M'_{\nu D})$ and $(M'_e, M'_{\nu D}, M'_\nu)$ to be physically equivalent, leading to the same mixing. However, in general, this may not hold for a more general $M_R$ structure with three different eigenvalues leading to a greater complexity in the structure of the light neutrino mass matrix, which may also deviate from the symmetric mass structure. *Note that the matrices $M'_e$ and $M'_{\nu D}$ are texture based* [10] *with no such restriction on $M_R$ and hence on $M'_\nu$.* It therefore becomes desirable to investigate the implications of the elements $(M')_{11,22}$ in the matrices $(M'_e, M'_{\nu D})$ for neutrino oscillation data, especially if the condition of naturalness of Eq. (2) is imposed on these.

## MATRIX DIAGONALIZATION

It may be noted that for the WB textures of lepton mass matrices mentioned in Eqs. (4) and (5), one of the lepton mass matrices is a Fritzsch-like texture two zero type [9] ($e_L = 0$) whereas the other has the following form

$$M'_L = \begin{pmatrix} e_L & |a_L|e^{i\alpha_L} & 0 \\ |a_L|e^{-i\alpha_L} & d_L & |b_L|e^{i\beta_L} \\ 0 & |b_L|e^{-i\beta_L} & c_L \end{pmatrix}, \quad L = e, \nu. \quad (6)$$

In order to construct the PMNS matrix, one needs to obtain the diagonalizing transformations for this matrix. It is observed [13,17,18] that for $|a_L|$ and $|b_L|$ to remain real, the free parameters get constrained within the limits

$$m_1 > e_L > -m_2 \text{ and } (m_3 - m_2 - e_L) > d_L > (m_1 - m_2 - e_L). \quad (7)$$

The above constraints indicate that the condition of Hermicity on the texture one zero mass matrix in Eq. (6) restrict the free parameters $e_L$ to have small values only, consistent with the naturalness condition in Eq. (2). The exact diagonalizing matrix $O_e$ for $M'_e$ defined through $D_e = O_e^\dagger P_e M'_e Q_e O_e$ is expressed as [18]

$$O_e = \begin{pmatrix} 1 & \sqrt{\frac{m_{e\mu}(1-\xi_e)}{(1+m_{\mu\tau})}} & \sqrt{\frac{m_{e\mu}m_{\mu\tau}(\zeta_e+m_{\mu\tau})(1-\xi_e)}{(1+m_{\mu\tau})}} \\ \sqrt{\frac{m_{e\mu}(1-\xi_e)}{(1+\zeta_e)}} & -\sqrt{\frac{1}{(1+\zeta_e)(1+m_{\mu\tau})}} & \sqrt{\frac{(\zeta_e+m_{\mu\tau})}{(1+\zeta_e)(1+m_{\mu\tau})}} \\ -\sqrt{\frac{m_{e\mu}(\zeta_e+m_{\mu\tau})(1-\xi_e)}{(1+\zeta_e)}} & \sqrt{\frac{(\zeta_e+m_{\mu\tau})}{(1+\zeta_e)(1+m_{\mu\tau})}} & \sqrt{\frac{1}{(1+\zeta_e)(1+m_{\mu\tau})}} \end{pmatrix}, \quad (8)$$

where $P_e = \text{diag}(e^{-i\alpha_e}, 1, e^{i\beta_e})$ with $Q_e = P_e^\dagger$ (for Hermitian $M'_e$) and $m_e \ll m_\mu$ along with $m_e \ll m_\tau$ have been used for the charged lepton masses. The free parameters $\xi_e$ and $\zeta_e$ represent the hierarchy characterizing parameters for the mass matrix $M'_e$ and are defined as $\xi_e = e_e/m_e$, $\zeta_e = d_e/c_e$ while $m_{e\mu} = m_e/m_\mu$, $m_{e\tau} = m_e/m_\tau$ along with $m_{\mu\tau} = m_\mu/m_\tau$ have been considered for simplicity.

The diagonalizing transformation $O_{\nu D}$ for matrix $M'_{\nu D}$, defined through

$$D_{\nu D} = O_{\nu D}^\dagger P_{\nu D} M'_{\nu D} Q_{\nu D} O_{\nu D} \quad (9)$$

is given by

$$O_{vD} = \begin{pmatrix} \sqrt{\dfrac{(1+\xi_{vD}m_{v12D})}{(1+m_{v12D})}} & \sqrt{\dfrac{m_{v12D}(1-\xi_{vD})}{(1+m_{v12D})(1+m_{v12D})}} & \kappa\sqrt{\dfrac{m_{v13D}m_{v23D}(m_{v23D}+\zeta_{vD})(1+\xi_{vD}m_{v12D})}{(1+m_{v23D})(1+m_{v23D})}} \\ \sqrt{\dfrac{m_{v12D}(1-\xi_{vD})(1+m_{v23D})}{(1+m_{v12D})(1+\zeta_{vD})}} & -\sqrt{\dfrac{(1+\xi_{vD}m_{v12D})}{(1+m_{v12D})(1+\zeta_{vD})}} & \sqrt{\dfrac{(m_{v23D}+\zeta_{vD})}{(1+m_{v23D})(1+\zeta_{vD})}} \\ -\sqrt{\dfrac{m_{v12D}(m_{v23D}+\zeta_{vD})}{(1+m_{v12D})(1+\zeta_{vD})}} & \sqrt{\dfrac{(1+\xi_{vD}m_{v12D})(m_{v23D}+\zeta_{vD})}{(1+m_{v12D})(1+m_{v23D})(1+\zeta_{vD})}} & \sqrt{\dfrac{1}{(1+\zeta_{vD})}} \end{pmatrix},$$

(10)

where $P_{vD} = \mathrm{diag}(e^{-i\alpha_{vD}}, 1, e^{i\beta_{vD}})$ with $Q_{vD} = P_{vD}^{\dagger}$ (for Hermitian $M'_{vD}$), $\kappa = \sqrt{(1-\xi_{vD})/(1-m_{v13})}$ and the free parameters $\xi_{vD}$, $\zeta_{vD}$ characterize the hierarchy that is exhibited by the elements of the Dirac neutrino mass matrix and are defined as $\xi_{vD} = e_{vD}/m_{v1D}$, $\zeta_{vD} = d_{vD}/c_{vD}$ while $m_{v12D} = m_{v1D}/m_{v2D}$, $m_{v13D} = m_{v1D}/m_{v3D}$ and $m_{v23D} = m_{v2D}/m_{v3D}$ have again been considered for simplicity. The condition of naturalness is imposed on the lepton mass matrices $(M'_e, M'_{vD})$ by restricting the parameter spaces of the free parameters to $(\zeta_e, \zeta_{vD}, \xi_e, \xi_{vD}) < 1$ and assuming NH for the Dirac neutrinos i.e. $m_{v1D} < m_{v2D} < m_{v3D}$.

## SEESAW AND PMNS MATRIX

Noting that the seesaw framework involves more free parameters than can be obtained from low energy data, it would be desirable to keep the structure of $M_R$ as simple as possible involving least number of free parameters. In this context, following FSTY, we have considered $M_R = m_R I$. In such a case, it can be shown [10,16] that the light neutrino mass matrix takes the following form

$$M'_v = -M'^T_{vD} M_R^{-1} M'_{vD} = P_{vD} O_{vD} \dfrac{(D_{vD})^2}{m_R} O^T_{vD} P_{vD}$$ (11)
$$= P_{vD} O_{vD} D_v O^T_{vD} P_{vD},$$

where $D_v = \mathrm{diag}(m_{v1}, -m_{v2}, m_{v3})$, so that the assumption of a NH for Dirac neutrinos given by $m_{v1D} < m_{v2D} < m_{v3D}$ automatically translates into a NH for light neutrino masses i.e. $m_{v1} < m_{v2} < m_{v3}$, and hence for the neutrino matrix $M'_v$. One can now easily compute the PMNS matrix [10,16] as

$$V = O_e^{\dagger} Q_e P_{vD} O_{vD}. \quad (12)$$

In order that the mixing angles are independent of $m_R$, we consider the neutrino masses $m_{v1}$, $m_{v2}$ and $m_{v3}$ as input parameters, so that we make the replacement of $m_{vD}$ with $\sqrt{m_v m_R}$ in all the terms of $O_{vD}$ in agreement with Eq. (11) above [10,16]. In such a case the ratios $m_{v12D}, m_{v13D}$ and $m_{v23D}$ get replaced by $m_{v12D} \to \sqrt{m_{v12}} = \sqrt{m_{v1}/m_{v2}}$ and so on. Likewise, the hierarchy parameters get redefined as $\xi_{vD} \to \xi_v = e_{vD}/\sqrt{m_v m_R} = e_v/\sqrt{m_{v1}}$ and $\zeta_{vD} \to \zeta_v = \zeta_{vD}$, where $\xi_v < 1$ and $\zeta_v < 1$, are again arbitrary and in accordance with the condition of naturalness. For the lepton mass matrices $(M'_e, M'_{vD}, M'_v)$ given by the Eqs. (4, 5, 11, 12), the product $Q_e P_{vD}$ in eqn. (11), is a diagonal phase matrix given by $Q_e P_{vD} = \mathrm{diag}(e^{-i\phi_1}, 1, e^{i\phi_2})$ [16], where the phases $\phi_1 = \alpha_e - \alpha_{vD}$ and $\phi_2 = \beta_e - \beta_{vD}$ are also free parameters.

### A. CASE I

For WB representation $(M'_e, M'_{vD})$ given in Eq. (4), the three lepton mixing angles may be expressed in terms of the lepton mass ratios $m_{e\mu}, m_{e\tau}, m_{\mu\tau}, \sqrt{m_{v12}}, \sqrt{m_{v13}}, \sqrt{m_{v23}}$, the free parameters $\zeta_e, \zeta_v, \xi_e$ and the phases $\phi_1$ and $\phi_2$ as

$$s_{12} = \sqrt{\dfrac{\sqrt{m_{v12}}}{\left(1+\sqrt{m_{v12}}\right)\left(1+\sqrt{m_{v23}}\right)}}, \quad (13)$$

$$s_{13} = \left| \sqrt{\dfrac{\sqrt{m_{v13}}\sqrt{m_{v23}}\left(\sqrt{m_{v23}}+\zeta_v\right)}{\left(1+\sqrt{m_{v23}}\right)\left(1-\sqrt{m_{v13}}\right)\left(1+\sqrt{m_{v23}}\right)}} e^{-i\phi_1} - \sqrt{\dfrac{m_{e\mu}(1-\xi_e)}{(1+\zeta_e)(1+\sqrt{m_{v23}})(1+\zeta_v)}} \left(\sqrt{\left(\sqrt{m_{v23}}+\zeta_v\right)} - \sqrt{\left(\zeta_e+m_{\mu\tau}\right)\left(\sqrt{m_{v23}}+1\right)} e^{i\phi_2}\right) \right|, \quad (14)$$

$$s_{23} = \left| \sqrt{\dfrac{1}{(1+\zeta_e)(1+m_{\mu\tau})(1+\sqrt{m_{v23}})(1+\zeta_v)}} \left(\sqrt{\left(\sqrt{m_{v23}}+\zeta_v\right)} - \sqrt{\left(\zeta_e+m_{\mu\tau}\right)\left(\sqrt{m_{v23}}+1\right)} e^{i\phi_2}\right) \right| \quad (15)$$

where only the leading order term (first) as well as the next to leading order terms have been retained in the expressions. It is observed that the above relations hold good within an error of less than a percent. It is noteworthy that $s_{12}$ depends predominantly on the neutrino mass ratio $\sqrt{m_{v12}}$. Likewise, it is also observed that $s_{23}$ is independent of $\xi_e$. This is easy to interpret as the parameter $\xi_e$ does not invoke any mixing among the second and the third generations of leptons. As a result, it should be interesting to investigate the implications of $\xi_e$, if any, for $s_{13}$ as well as those of $\zeta_e$ and $\zeta_v$ for $s_{13}$ and $s_{23}$.

### B. CASE II

For the WB representation $(M'_e, M'_{vD})$ given in eqn. (5), the mixing angles may be expressed as

$$s_{12} = \sqrt{\dfrac{\sqrt{m_{v12}}(1-\xi_v)}{\left(1+\sqrt{m_{v12}}\right)\left(1+\sqrt{m_{v23}}\right)}}, \quad (16)$$

$$s_{13} = \left| \sqrt{\dfrac{\sqrt{m_{v13}}\sqrt{m_{v23}}\left(\sqrt{m_{v23}}+\zeta_v\right)\left(1+\xi_v\sqrt{m_{v12}}\right)(1-\xi_v)}{\left(1+\sqrt{m_{v23}}\right)\left(1-\sqrt{m_{v13}}\right)\left(1+\sqrt{m_{v23}}\right)}} e^{-i\phi_1} - \sqrt{\dfrac{m_{e\mu}}{(1+\zeta_e)(1+\sqrt{m_{v23}})(1+\zeta_v)}} \left(\sqrt{\left(\sqrt{m_{v23}}+\zeta_v\right)} - \sqrt{\left(\zeta_e+m_{\mu\tau}\right)\left(1+\sqrt{m_{v23}}\right)} e^{i\phi_2}\right) \right|, \quad (17)$$

$$s_{23} = \left| \sqrt{\frac{1}{(1+\zeta_e)(1+m_{l\pi})(1+\sqrt{m_{v23}})(1+\zeta_v)}} \left( \sqrt{(\sqrt{m_{v23}}+\zeta_v)} - \sqrt{(\zeta_e+m_{l\pi})(1+\sqrt{m_{v23}})} e^{i\phi_2} \right) \right|.$$

(18)

One observes that the mixing angle $(s_{12})^2$ depends predominantly on the neutrino mass ratio $\sqrt{m_{v12}}$ and the parameter $\xi_v$. As expected, the mixing angle $s_{23}$ is still independent of the parameter $\xi_v$. As a result, it should be interesting to investigate the implications of $\xi_v$, if any, for $s_{12}$ and $s_{13}$ as well as those of $\zeta_e$ and $\zeta_v$ for $s_{13}$ and $s_{23}$.

## INPUTS

We use the following recent global three neutrino oscillation data [19] at 1σ C.L. as inputs for our analysis.

$\delta m^2 = (7.32 - 7.80) \times 10^{-5}$ GeV$^2$,
$\Delta m^2 = (2.33 - 2.49) \times 10^{-3}$ GeV$^2$,
$\sin^2 \theta_{12} = 0.29 - 0.33$,
$\sin^2 \theta_{13} = 0.022 - 0.027$,
$\sin^2 \theta_{23} = 0.37 - 0.41$, (19)

where the neutrino mass square differences are defined as $\delta m^2 = m_{v2}^2 - m_{v1}^2$ and $\Delta m^2 = m_{v3}^2 - (m_{v1}^2 + m_{v2}^2)/2$ for NH [19]. The lightest neutrino mass $m_{v1}$, the parameters $\zeta_e, \zeta_v, \xi_e, \xi_v$ and phases $\phi_1$ and $\phi_2$ are taken to be free parameters. It is noteworthy that the neutrino mass ratios are completely determined if $m_{v1}$ is known, since

$$m_{v12} = \sqrt{\frac{m_{v1}^2}{m_{v1}^2 + \delta m^2}} \text{ and } m_{v13} = \sqrt{\frac{m_{v1}^2}{m_{v1}^2 + \Delta m^2 + (\delta m^2)/2}}. \quad (20)$$

In addition, we have imposed the condition of naturalness on the lepton mass matrices through the constraints $(\zeta_e, \zeta_v, \xi_e, \xi_v) < 1$ and assumed NH for the neutrino masses, consistent with the condition of naturalness. Furthermore, in the absence of any clues for CP violation in the lepton sector, the phases $\phi_1$ and $\phi_2$ have been given full variation from 0 to $2\pi$.

## RESULTS

### A. CASE I

It is observed that the complete 1σ range of all the neutrino oscillation parameters given in eqn. (19) can be reconstructed by the relations (13-15). Interestingly, it is observed that all values for $\zeta_e$, $\xi_e$ allowed by the condition of naturalness i.e. $0 < (\xi_e, \zeta_e) < 1$ do not play any significant role in fixing the mixing angles which is depicted in **Fig. 1**, with respect to $\xi_e$. Similar plot is obtained with respect to $\zeta_e$. This is easily understandable for $s_{12}$ and $s_{23}$ through Eqs. (13, 15) with respect to $\xi_e$. However for $s_{13}$, it is noticed that nonzero values of $\zeta_e$ and $\xi_e$ appear only in the next to leading order term in Eq. (14), the contribution of which is very small as compared to the corresponding leading order term, as a result the implications of nonzero $\zeta_e$ and $\xi_e$ on $s_{13}$ are largely insignificant. For the mixing angle $s_{23}$, the contributions of the free parameter $\zeta_v$ are able to compensate for the nonzero values of $\zeta_e$ allowed by the condition of naturalness, without affecting the predicted values of $s_{23}$ as shown in Fig. 2.

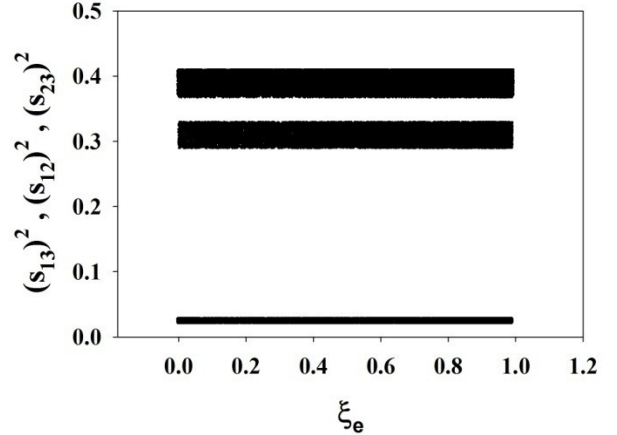

**Fig. 1: Plot showing redundancy of the parameter $\xi_e$ in estimating the lepton mixing angles for case I.**

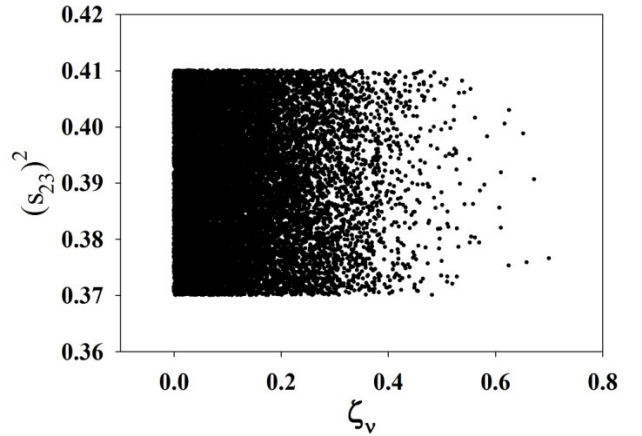

**Fig. 2: Plot showing redundancy of parameter $\zeta_v$ for $(s_{23})^2$ in the case I.**

Since the unique window to verify the Majorana nature of massive neutrinos is through the neutrinoless double beta (0νββ) decay, it is also desirable to study the impact of the parameters $\zeta_e$ and $\xi_e$ on the effective mass $m_{ee}$ measured in (0νββ) defined through [20]

$$m_{ee} = m_{v1} |V_{e1}|^2 + m_{v2} |V_{e2}|^2 + m_{v3} |V_{e3}|^2. \quad (21)$$

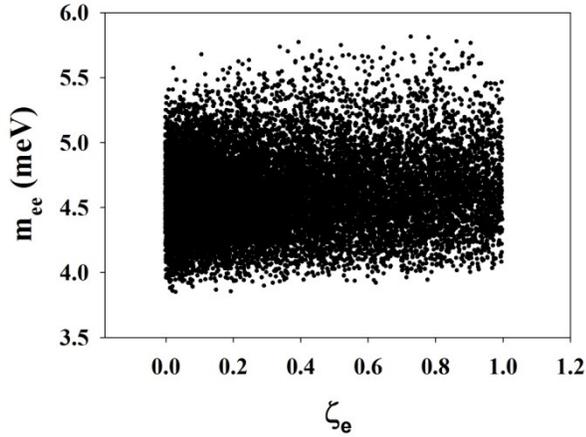

Fig. 3: Plot depicting that the parameter $\zeta_e$ does not contribute in determination of $m_{ee}$ for case I.

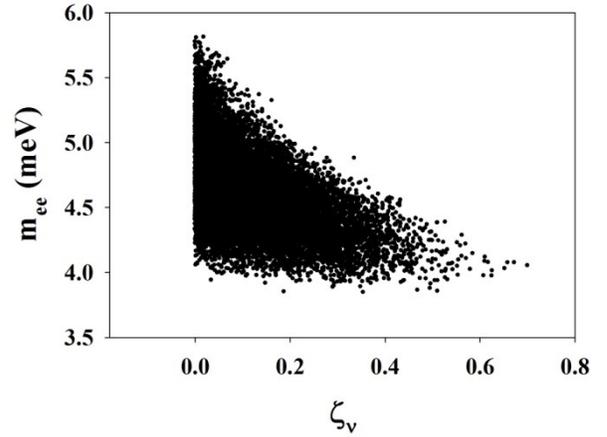

Fig. 5: Plot indicating the strong influence of the nonzero values of the parameter $\zeta_v$ on $m_{ee}$ for case I.

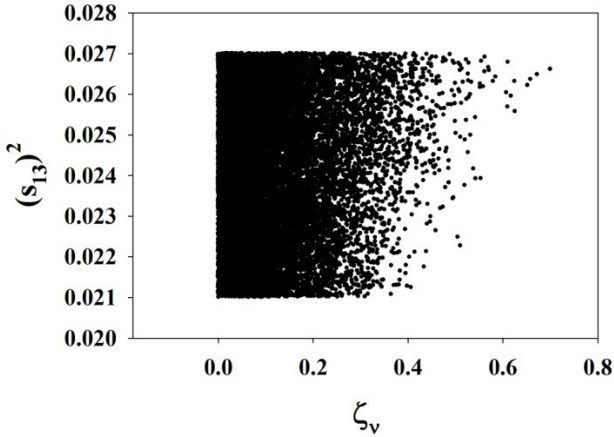

Fig. 4: Plot showing the dependence of the mixing angle $(s_{13})^2$ on the parameter $\zeta_v$ for case I.

Interestingly, we find that nonzero $\zeta_e$ and $\xi_e$ do not contribute in determining $m_{ee}$, as depicted in **Fig. 3** for $\zeta_e$. Similar observations are also obtained for $\xi_e$ indicating that $\xi_e$ and $\zeta_e$ may most generally be considered to be redundant in neutrino oscillations. However, from the Eq. (14) it is noticed that due to the presence of $\left(\sqrt{m_{v23}}+\zeta_v\right)$ in the leading order term for $s_{13}$, the phenomenologically allowed range of $\zeta_v$ is limited from $0<\zeta_v<1$ to $0<\zeta_v<0.70$ in order to regenerate the experimentally measured $s_{13}$, as illustrated in the **Fig. 4**, wherein it is also noticed that values of $\zeta_v > 0.70$ lead to an overshoot in $s_{13}$ from its experimental range. On one hand, one is able to regenerate the complete $1\sigma$ range of all the three mixing angles for $0 \leq \zeta_v < 0.7$, indicating that the parameter $\zeta_v$, like $\zeta_e$ and $\xi_e$, also has no significant impact on neutrino mixing angles yielding corresponding graphs similar to **Fig. 1**.

On the other hand, interesting observations are obtained for $m_{ee}$. The large values of the parameter $\zeta_v$ appear to have greater implications on the allowed range of $m_{ee}$, as shown in **Fig. 5**. On the contrary, all the allowed values of the remaining parameters $\xi_e$ and $\zeta_e$ including zero have no impact on $m_{ee}$ signifying that the nonzero values of $\zeta_v$ imply vital implications especially in the context of $m_{ee}$ and hence only the parameters $\xi_e$ and $\zeta_e$ may be considered to be completely redundant if neutrinos are Majorana particles leading to Fritzsch-like texture five zero structures of the corresponding lepton mass matrices i.e. with ($\zeta_e = 0$, $\xi_e = 0$, $\xi_v = 0$, $\zeta_v \neq 0$).

### B. CASE II

Interestingly, the complete $1\sigma$ range of all the neutrino oscillation parameters given in Eqn. (20) can also be regenerated by the relations (16-18). However, using the Eqs. (16, 17), it is noticed that due to the term $\sqrt{(1-\xi_v)}$, the phenomenologically allowed range of $\xi_v$ is limited from $0<\xi_v<1$ to $0<\xi_v<0.50$ in order to regenerate the experimentally measured $s_{12}$ and $s_{13}$ and that the nonzero values of $\xi_v$ do not play any significant role in determining the mixing angles. This is obvious from Eqs. (16, 18) for $s_{12}$ and $s_{23}$. However, in the case of Eq. (17), the effect of increase in $\xi_v$ on $\sqrt{(1-\xi_v)}$ in the leading order term is compensated by a corresponding increase in $\sqrt{(1+\xi_v m_{v12})}$ with no significant impact on $s_{13}$. Furthermore, the principle conclusions drawn from the **Fig. 2, Fig. 3 and Fig. 4** in the case I are also obeyed for this case due to similar reasoning, indicating that the parameters $\zeta_v$, $\zeta_e$ and $\xi_v$ have no observed implications for neutrino mixing angles.

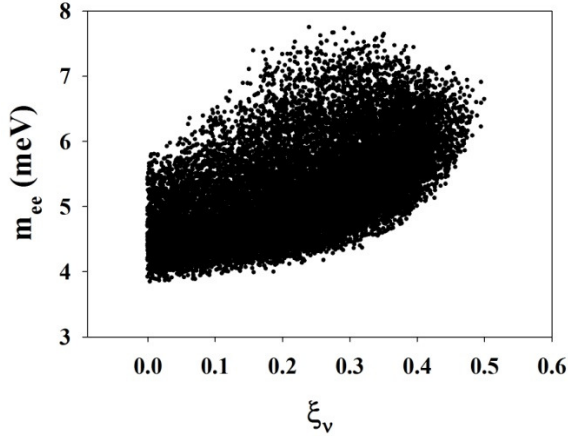

**Fig. 6:** Plot showing influence of nonzero values of the parameter $\xi_\nu$ on $m_{ee}$ for case II.

Although the allowed values of the parameter $\zeta_e$ continue to have no impact on $m_{ee}$, for this case, the same is not observed to be true for the remaining parameters namely $\xi_\nu$ and $\zeta_\nu$. The plot depicting the dependence of $m_{ee}$ on $\xi_\nu$ is shown in **Fig. 6**, while that for $m_{ee}$ versus $\zeta_\nu$ is observed to be the similar to **Fig. 5**, reinforcing that the nonzero values of the parameters $\xi_\nu$ and $\zeta_\nu$ have vital implications on the allowed range of $m_{ee}$. Interestingly, both the non-redundant parameters $\xi_\nu$ and $\zeta_\nu$ are involved in the neutrino mass matrix. As a result only the parameter $\zeta_e$ may be considered to be completely redundant for this case, leading to four texture zeros involved in the resulting lepton mass matrices, i.e. with ($\zeta_e = 0$, $\xi_e = 0$, $\xi_\nu \neq 0$, $\zeta_\nu \neq 0$).

## CONCLUSIONS

We observe that even though the neutrino mixing pattern is significantly different from the quarks mixing pattern, yet it can also be described by *natural* mass matrices. Using the considerations of unitary transformations, seesaw mechanism and a hierarchical parameterization for the lepton mass matrices, through the condition of naturalness i.e. $(\zeta_e, \zeta_\nu, \xi_e, \xi_\nu) < 1$, we have been able to illustrate the effect of lepton mass hierarchies on lepton mass mixing by exact relations involving lepton mass ratios, the hierarchy characterizing parameters and the phases $\phi_1$ and $\phi_2$. It has been clearly shown, that for the recent three neutrino oscillation data at 1σ C.L., the most general texture three zero lepton mass matrices of Eqs. (4, 5), obtained through unitary WB transformations, are physically equivalent to texture six zero Hermitian lepton mass matrices with $\xi_e = 0$, $\xi_\nu = 0$, $\zeta_e = 0$ and $\zeta_\nu = 0$, when the condition of naturalness is imposed on these and considerations of $m_{ee}$ are not taken into account. In particular, we find that no additional restrictions are imposed on the three mixing angles by the inclusion of non-zero values of the additional free parameters $\xi_e$, $\xi_\nu$, $\zeta_e$ and $\zeta_\nu$. For such texture six zero lepton mass matrices, the relations for the three mixing angles reduce to the ones derived by FSTY [10], if one ignores the contributions of $\sqrt{m_{\nu 12}}$, $\sqrt{m_{\nu 23}}$, $\sqrt{m_{\nu 23}}$ and $m_{\mu\tau}$ as compared to unity and considers $(m_{\nu 23})^{1/4} > \sqrt{m_{\mu\tau}}$.

On the contrary, if $m_{ee}$ is taken into consideration, Fritzsch-like texture five zeros with $\xi_e = 0$, $\xi_\nu = 0$, $\zeta_e = 0$ and $\zeta_\nu \neq 0$ naturally emerge as the most general (or generic) lepton mass textures involving a maximum number of texture zeros. It is perhaps desirable to study the implications of these texture five zero structures for the possible CP violation in the lepton sector. In view of this, we investigate, in Fig. 7, the implications of the non-redundant free parameter $\zeta_\nu$ for the Jarlskog's rephasing invariant CP-violation measure $J_{CP}$ [21] defined through

$$J_{CP} = \text{Im}\left[V_{\mu 3}V^*_{\tau 3}V^*_{\mu 2}V_{\tau 2}\right]. \quad (22)$$

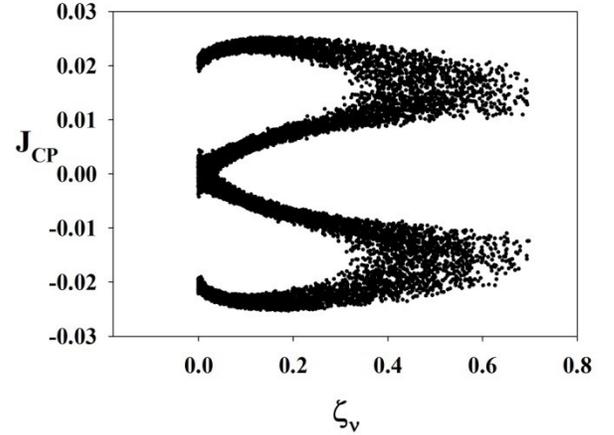

**Fig. 7:** Plot showing the implications of the non-redundant parameter $\zeta_\nu$ for $J_{CP}$.

We find that the current three neutrino oscillation data at 1σ in Eqn. (20) allow $J_{CP} = 0$ for $0 \leq \zeta_\nu < 0.045$, but the higher non-zero values of $\zeta_\nu$ have vital implications on the allowed values of $J_{CP}$ i.e. $-0.025 < J_{CP} < 0.025$, thereby reinforcing the non-redundancy of $\zeta_\nu$ for neutrino oscillation phenomenology as well as indicating towards a possible CP violation in the lepton sector, in agreement with Ref. [10]. In particular, for such matrices with $\xi_e = 0$, $\xi_\nu = 0$, $\zeta_e = 0$ and $\zeta_\nu \neq 0$, we obtain the following limits for $m_{ee}$, $m_{\nu 1}$, $m_{\nu 2}$ and $m_{\nu 3}$, e.g.

$$3.7 \text{ meV} < m_{ee} < 5.7 \text{ meV},$$
$$0.3 \text{ meV} < m_{\nu 1} < 2.0 \text{ meV},$$
$$8.5 \text{ meV} < m_{\nu 2} < 9.0 \text{ meV},$$

$$48.6 \text{ meV} < m_{\nu 3} < 50.3 \text{ meV}, \quad (23)$$

which are in good agreement with some of the recent analysis by several authors [7, 10, 16, 22]. These neutrino masses are stable to the radiative corrections introduced by the Renormalization Group Equations [23] when the lepton mass matrices are energy rescaled. It is observed that for $m_R \sim O(10^{10})$ GeV, these corrections are several orders of magnitude ($10^{-6}$ to $10^{-5}$ times) smaller [10] than the neutrino masses in Eqn. (23). Such corrections do not appear to disturb the natural structure of the corresponding lepton mass matrices and hence the results obtained in this paper are unaffected by these corrections. Furthermore, the above values of neutrino masses appear to favor standard leptogenesis as the mechanism to produce the Baryon Asymmetry in the Universe [24].

## ACKNOWLEDGEMENT

The author would like to thank the Director, Rayat Institute of Engineering and Information Technology, for providing the necessary working facilities.